\documentclass[prl,twocolumn,floatfix]{revtex4}

\usepackage{graphicx,amssymb,amsmath}

\begin{document}

\title{High Intrinsic Mobility and Ultrafast Carrier Dynamics in Multilayer Metal Dichalcogenide MoS$_2$}

\author{Jared H. Strait}
\email{jhs295@cornell.edu}
\author{Parinita Nene}
\author{Farhan Rana}
\affiliation{School of Electrical and Computer Engineering, Cornell University, Ithaca, NY, USA}

\begin{abstract}
The ultimate limitations on carrier mobilities in metal dichalcogenides, and the dynamics associated with carrier relaxation, are unclear.  We present measurements of the frequency-dependent conductivity of multilayer dichalcogenide MoS$_2$ by optical-pump terahertz-probe spectroscopy.  We find mobilities in this material approaching 4200 cm$^2$V$^{-1}$s$^{-1}$ at low temperatures.  The temperature dependence of scattering indicates that the mobility, an order of magnitude larger than previously reported for MoS$_2$, is intrinsically limited by acoustic phonon scattering at THz frequencies.  Our measurements of carrier relaxation reveal picosecond cooling times followed by recombination lasting tens of nanoseconds and dominated by Auger scattering into defects.  Our results provide a useful context in which to understand and evaluate the performance of MoS$_2$-based electronic and optoelectronic devices.
\end{abstract}

\maketitle
Layered two-dimensional transition metal dichalcogenides have recently enjoyed a resurgence of interest from the scientific community both from a new science perspective and also for novel applications\cite{Podzorov04,Splendiani10,Mak10,Eda11,Yoon11,Cao12,Kioseoglou12,VanDerZande13,Zhou13,Wang12}. In contrast to graphene, metal dichalcogenides have non-zero bandgaps and are also efficient light emitters, making them attractive for electronics and optoelectronics\cite{Podzorov04,Splendiani10,Mak10,Eda11,Yoon11,Cao12,Kioseoglou12,VanDerZande13,Zhou13,Wang12}. The ability to synthesize atomically thin semiconducting crystals and their heterostructures and transfer them to arbitrary substrates has opened the possibility of transparent, flexible electronics and optoelectronics based on these material systems\cite{Wang12,WangH12,Laskar13,Lopez13,Perkins13,Malard13,Zhu14,Mak14}.  The best reported carrier mobilities in metal dichalcogenides, typically in the few hundred cm$^{2}$V$^{-1}$s$^{-1}$ range for MoS$_{2}$\cite{Das12,Bao13}, are not as large as in graphene.  But the reported mobilities are large in comparison to those of organic materials often used for flexible electronics\cite{Podzorov04}. Physical mechanisms limiting the mobility of monolayer and multilayer MoS$_2$ field effect transistors have been the subject of several experimental\cite{Fivaz67,Das12,Kim12,Bao13,Pradhan13,Radisavljevic13,Kumar13} and theoretical\cite{Kaasbjerg12,Ma14} investigations. Charged impurity scattering, electron-phonon interaction, and screening by the surrounding dielectric environment are all believed to affect the mobility. Due to the challenge of isolating these effects, the intrinsic mobility of MoS$_2$, and the ultimate performance of electronic devices, remains unclear. Similarly, carrier intraband and interband scattering and relaxation rates, which determine the performance of almost all proposed and demonstrated electronic and optoelectronic metal dichalcogenide devices, remain poorly understood.   

\begin{figure}[tbp]
	\centering
		\includegraphics[width=.47\textwidth]{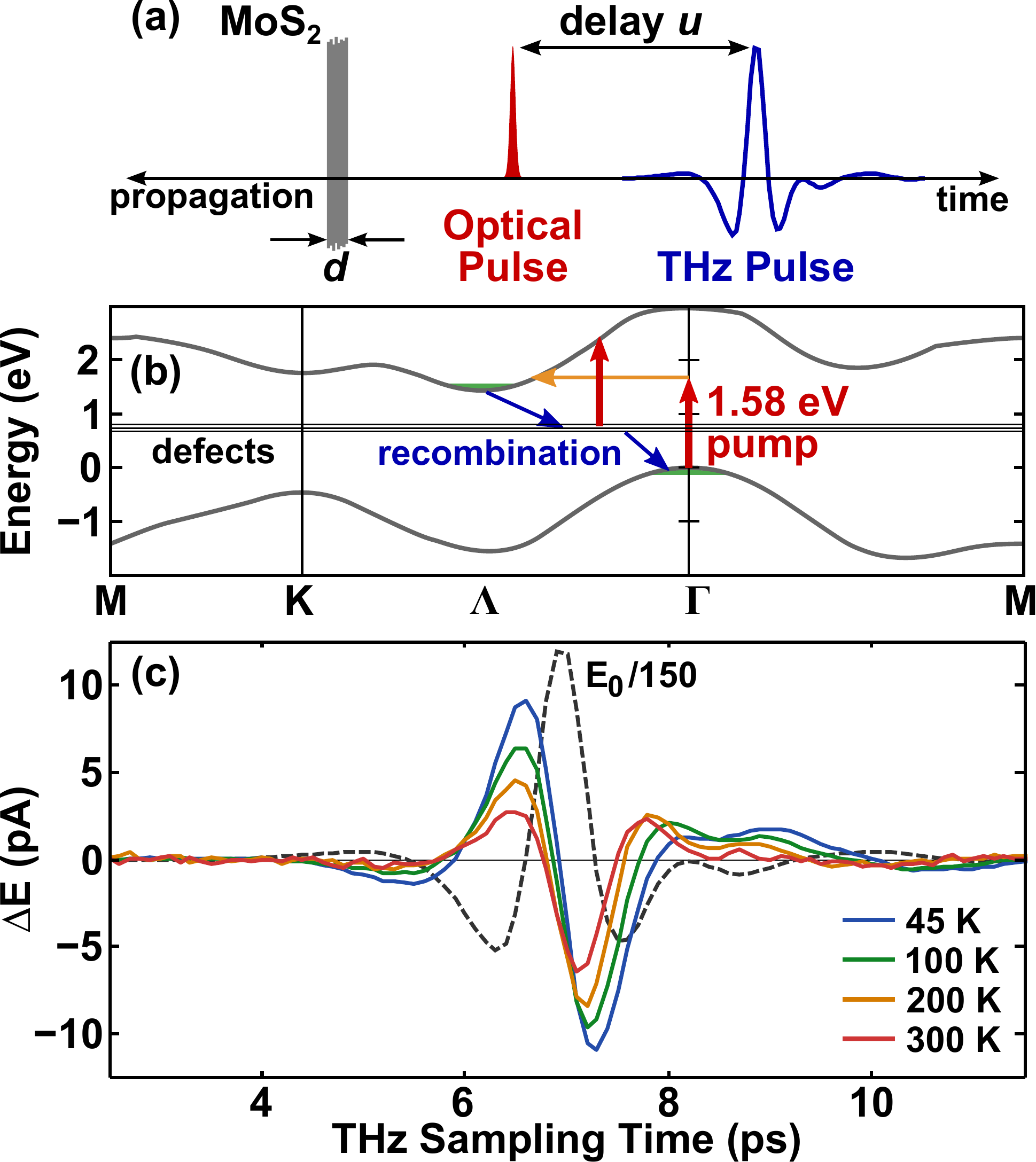}
	\caption{Optical-Pump THz-Probe Spectroscopy. (a) A schematic of optical-pump and THz-probe measurement scheme. (b) In-plane bandstructure of multilayer MoS$_{2}$ depicting carrier photoexcitation and recombination.  (c) (Solid) Measured $\Delta E(t,u)$ in units of detector current for a fixed pump-probe delay $u=5$ ps.  $\Delta E(t,u=-5$ ps) is subtracted for clarity.  (Dashed) The scaled reference THz pulse $E_0(t)$.
	}
	\label{fig:DeltaE}
\end{figure}

In this report, we present optical-pump THz-probe measurements of the time- and frequency-dependent conductivity of multilayer MoS$_2$. Previously reported electrical measurements\cite{Das12,Kim12,Bao13,Pradhan13} have probed the MoS$_2$ DC carrier transport, and all-optical measurements\cite{Korn11,Kumar13} have probed the exciton temporal dynamics.  In contrast, optical-pump THz-probe spectroscopy can measure the time development of the complex intraband conductivity in the 0.2-2.0 THz frequency range with a temporal resolution better than a few hundred femtoseconds.  The frequency dependence of the measured intraband conductivity leads to the direct extraction of the carrier momentum scattering rate, from which the carrier mobility can be determined in a contact-free way.  In addition, the temporal development of the intraband conductivity reveals carrier interband and intraband relaxation dynamics.

Our results show that the momentum scattering times in multilayer MoS$_2$ vary from $\sim$90 fs at 300 K to $\sim$1.46 ps at 30 K, corresponding to carrier mobilities of $\sim$257 and $\sim$4200 cm$^{2}$V$^{-1}$s$^{-1}$, respectively. The temperature dependence of the momentum scattering rate reveals that the mobility is limited by acoustic phonon scattering and the measured scattering times agree well with the theoretical predictions\cite{Kaasbjerg12,Ma14}. The measured mobility values are almost an order of magnitude larger than the previously reported values in multilayer MoS$_2$ devices\cite{Fivaz67,Kim12,Bao13}. Our results indicate that the mobilities in reported MoS$_2$ electronic devices\cite{Fivaz67,Das12,Kim12,Pradhan13,Radisavljevic13} are  likely limited by extrinsic mechanisms (defects, impurities, grain boundaries, etc.) and point to the ultimate performance of metal dichalcogenide electronic devices. Our measurements of carrier relaxation reveal picosecond cooling times varying from $\sim$0.7 ps at 300 K to $\sim$1.2 ps at 45 K.  Subsequent density-dependent recombination of carriers occurs over 10's of nanoseconds at low temperature.  We present a recombination model based on carrier capture into defect states which reproduces the measured nonlinearity of this recombination versus carrier density as well as the observed temporal dynamics.

\section{Results}

\subsection{Conductivity spectra and momentum scattering times}
THz-frequency radiation is sensitive to the frequency-dependent intraband conductivity of MoS$_2$, $\sigma(\omega)$, which contains information about carrier scattering and mobility.  Using a THz time-domain spectrometer\cite{Katzenellenbogen91}, we measure the amplitude and phase of broadband THz pulses transmitted through our 4 $\mu$m thick natural MoS$_2$ samples.  In a static transmission measurement, where the conductivity is time-independent, the thin-film ($\omega d/c\ll1$) amplitude transmission for THz pulses is (see supplementary information),
\begin{equation}
	t(\omega)\approx\left[ 1-i\frac{\omega d}{2c} (n^{2}-1) + \frac{\eta_0\sigma(\omega) d}{2} \right]^{-1} \label{eq:tr}
\end{equation}
where $\omega$ is the frequency, $d$ is the sample thickness, $c$ is the speed of light, $\eta_0$ is the impedance of free space, $n$ is the sample refractive index.  Using Equation \ref{eq:tr} and the measured spectra of THz pulses transmitted through an unpumped multilayer MoS$_2$ sample, we obtained a constant value of refractive index $n\approx3.0$ in the 0.2-2.0 THz frequency range.  Further, the unpumped DC conductivity $\sigma_{0}$ was found to be in the range 10-20 S m$^{-1}$ indicating that the sample was very mildly doped.  Electrical measurements on similar single-layer and multilayer samples have indicated that these MoS$_{2}$ samples tend to be n-doped\cite{Changjian14}. Given the large value of the refractive index and the small value of $\sigma_{0}$, this unpumped transmission measurement could not reliably determine the frequency dependence of the conductivity and thereby the mobility. In contrast, since the change in THz transmission, $\Delta E(\omega)$, between pumped and unpumped samples is directly proportional to the change in the conductivity $\Delta\sigma(\omega)$, optical-pump THz-probe spectroscopy is a highly-sensitive method to extract the frequency-dependent conductivity.

In the optical-pump THz-probe scheme, depicted in Figure \ref{fig:DeltaE}(a-b), 780 nm center wavelength (1.58 eV), sub-100 fs optical pulses excite electron and hole distributions at the six $\Lambda$ points and the $\Gamma$ point, respectively, after which the distributions cool and recombine. The field of the THz pulse transmitted through the pumped sample, $E_{\rm p}(t,u)$, is sensitive to the evolving conductivity and, therefore, depends on the pump-probe delay $u$.  In our experiments, the incident optical and THz pulses are mechanically chopped at different frequencies and the transmitted THz pulse is measured at the sum frequency. We therefore measure the change in the transmitted THz pulse, $\Delta E(t,u)=E_{\rm p}(t,u)-E_{0}(t)$, in the presence ($E_{\rm p}$) and absence ($E_0$) of optical pumping. In this paper, the symbol $\Delta$ signifies the change in a quantity due to optical pumping. Figure \ref{fig:DeltaE}(c) shows the measured $\Delta E(t,u)$ for a fixed delay of $u=5$ ps at different temperatures.  We also display $E_{0}(t)$, which was essentially temperature-independent.  $\Delta E(t,u)$ contains information of both the amplitude and phase changes of $E_{0}(t)$ upon pumping. Comparison of $\Delta E(t,u)$ and $E_{0}(t)$ shows a trend of increasing phase shifts with decreasing temperatures due to photoexcited carriers. Thus, both the amplitude and the phase of the transmitted THz pulse are needed to determine the sample response.

The sample response can be determined from the measured change in the transmitted THz pulse, $\Delta E(t,u)$, as follows. We scan the pump-probe delay simultaneously with the transmitted THz pulse in the time-domain THz spectrometer such that each measured point of the transmitted THz pulse is at the same delay, $u$, from the pump\cite{Kindt99,Beard00,Nienhuys05}. The measured pulse, $\Delta \tilde{E}(t,u) = \Delta E(t,u-t)$, satisfies (see supplementary information),
\begin{equation}
\Delta \tilde{J}(t,u) = - \frac{2}{\eta_0 d}\Delta \tilde{E}(t, u) - \epsilon_0 (n^2-1)\frac{\partial\Delta \tilde{E}(t, u)}{\partial t} \label{eq:DeltaJ}
\end{equation}
where $\epsilon_0$ is the permittivity of free-space. Equation \ref{eq:DeltaJ} describes the effect of the free-carrier and polarization current densities on the THz pulse. The change in the free-carrier current density, $\Delta \tilde{J}(t,u) = \Delta J(t,u-t)$, in the sample can be written approximately as (see supplementary information),
\begin{equation}
\Delta \tilde{J}(t,u) \approx E_0(t) \otimes [\Delta \sigma_{\rm DC}(u-t) j(t)] \label{eq:J}
\end{equation}
Here, $\Delta \sigma_{\rm DC}(t)$ is the change in the DC conductivity and $j(t)$ is the normalized current impulse response ($\int_{-\infty}^{\infty} dt j(t)=1$). In Fourier domain, we define $\Delta \sigma(\omega,u)$ as $\Delta \sigma_{\rm DC}(u) j(\omega)$. Techniques to extract $\Delta \sigma(\omega,u)$ from the measured $\Delta \tilde{E}(t,u)$ are described in the Methods section. 
\begin{figure}[tbp]
	\centering
		\includegraphics[width=.47\textwidth]{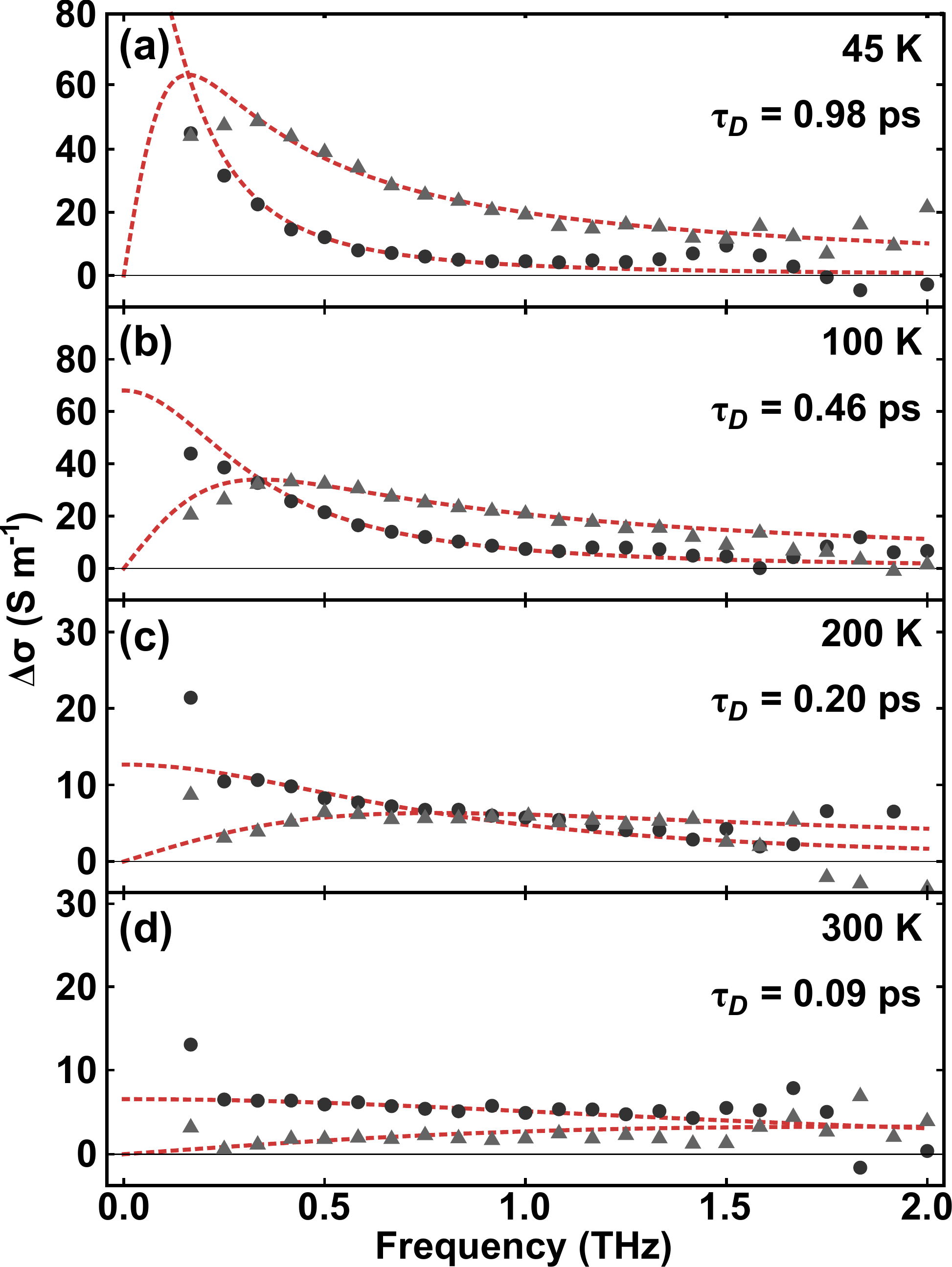}
	\caption{THz conductivity.  The real (circles) and imaginary (triangles) parts of the measured THz conductivity $\Delta\sigma(\omega,u)$ plotted versus frequency for various temperatures.  In each case, the probe delay is $u=5$ ps, and the pump fluence is $\sim$1.2 $\mu$J cm$^{-2}$. (Dashed lines) Drude model fits to the measured conductivity spectra used to extract $\tau_{\rm D}$.} 
	\label{fig:DrudeFits}
\end{figure}

Figure \ref{fig:DrudeFits} shows the real and imaginary parts of the photoexcited conductivity $\Delta\sigma(\omega,u)$ measured at four different temperatures in MoS$_2$. We find that the measured conductivity spectra closely follow the Drude form, $\Delta \sigma_{\rm DC}(u)/(1-i\omega\tau_{\rm D})$, which is the simplest conductivity model in the relaxation time approximation. Therefore, the current impulse response $j(\omega)$ is $(1-i\omega\tau_{\rm D})^{-1}$. We fit the real and imaginary parts of the data simultaneously with a weighted least-squares regression and extract the photoexcited DC conductivity $\Delta\sigma_{\rm DC}(u)$ and the carrier momentum scattering time $\tau_{\rm D}$.  Weights were the data variance at each frequency point over twenty scans.  The high quality of fits to the data obtained at all temperatures, as shown in Figure \ref{fig:DrudeFits}, is strong evidence of the fact that the change in THz transmission we measure indeed originates from the intraband conductivity of the photoexcited carriers.  Also, at all temperatures we find no significant variation in the extracted value of $\tau_{\rm D}$ for any value of $u$ in the range 5 ps $\le u \le$ 12 ns. Since one expects very hot carriers to have average momentum scattering times different from that of cold carriers\cite{Jnawali13}, this observation suggests that the carrier distributions after photoexcitation cool down on time scales shorter than a few picoseconds.  We further discuss carrier cooling times below.  

\subsection{Carrier mobility}

\begin{figure}[tbp]
	\centering
		\includegraphics[width=.47\textwidth]{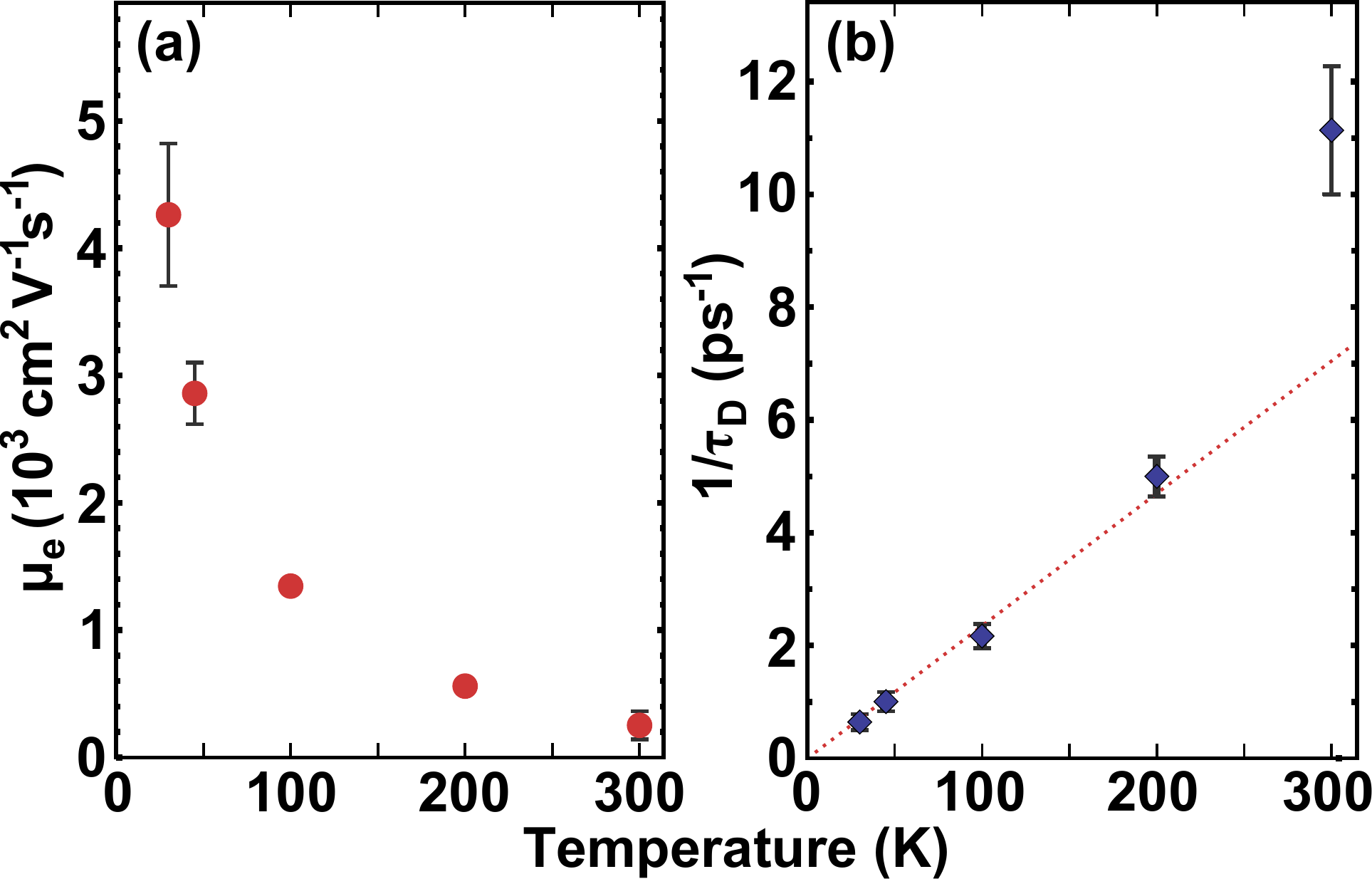}
	\caption{Mobility and scattering rate.  (a) Electron mobility versus temperature, determined from the measured momentum scattering rate.  (b) Measured momentum scattering rate versus temperature with a linear fit for $T \le 200$ K.  The $1/\tau_{D}\propto T$ dependence at low temperatures suggests acoustic phonon scattering as the dominant scattering mechanism.  Meascured mobility reaches $\sim$4200 cm$^2$V$^{-1}$s$^{-1}$ at 30 K.}
	\label{fig:Mobility}
\end{figure}

Carrier in-plane ($\perp$ to c-axis) mobility is related to the momentum scattering time by $\mu = e \tau_{\rm D}/m^*_{\rm c}$, where $m^*_{\rm c}$ is the in-plane conductivity effective mass.  Multilayer MoS$_{2}$ has six electron pockets in the Brillouin zone, each with an anisotropic effective mass tensor\cite{Peel12,Zahid13}.  The one hole pocket has an isotropic effective mass tensor.  Using DFT values for the electron and hole effective mass tensors\cite{Peel12}, we find the conductivity effective masses for both electrons and holes to be $\sim$0.61$m_{0}$. In Figure \ref{fig:Mobility}(a), we plot the electron mobilities corresponding to the measured momentum scattering times at different temperatures ($T$ = 30, 45, 100, 200, 300 K). We find a mobility of 257 cm$^2$V$^{-1}$s$^{-1}$ at 300 K, increasing to 4200 cm$^2$V$^{-1}$s$^{-1}$ at 30 K.  For 300 K, the measured mobility is consistent with recent electronic transport measurements in multilayer MoS$_2$\cite{Das12,Kim12,Bao13,Pradhan13,Fivaz67}.  But the mobilities we find below 200 K in this contact-free measurement are significantly higher than the values previously reported for MoS$_2$.

In Figure \ref{fig:Mobility}(b), we plot the carrier momentum scattering rate, $1/\tau_{\rm D}$, versus temperature.  The scattering rate for $T\le200$ K increases linearly with temperature.  This linear temperature dependence is expected for quasi-elastic acoustic phonon scattering in 2D and layered materials in the equipartition regime\cite{Fivaz67,Kawamura92,Kaasbjerg12,Ma14}.  Other scattering mechanisms, such as impurity scattering, have a different temperature dependence for layered materials\cite{Ma14}.  The larger scattering rate observed at the highest temperature (300 K) likely indicates an optical phonon scattering contribution\cite{Kaasbjerg12,Kim12}. In the deformation-potential approximation, the energy-independent acoustic-phonon-limited momentum scattering rate in 2D and layered materials is related to the temperature by\cite{Kaasbjerg12,Ma14},
\begin{equation}
\frac{1}{\tau_{\rm D}} = \frac{m^*_{\rm d}\Xi^2 k_{\rm B}T}{\hbar^3\rho v_{\rm s}^2}
\label{eq:DefP}
\end{equation}
where $\Xi$ is the deformation potential, $\rho$ is the 2D mass density, $v_{\rm s}$ is the LA phonon velocity, and $m^*_{\rm d}$ is the carrier density of states effective mass. From the data in Figure \ref{fig:Mobility}(b), we find $d\tau_{\rm D}^{-1}/dT = 22.6\pm1.5$ ns$^{-1}$K$^{-1}$ for $T \le 200$ K.  Using $v_{s}=6.7\times10^3$ m s$^{-1}$\cite{Kaasbjerg12}, $\rho=3\times10^{-6}$ kg m$^{-2}$, $m^*_{\rm d}=0.62m_{0}$ (ab-initio values for $m^*_{\rm d}$ are almost identical for electrons and holes\cite{Peel12}), we find  $\Xi=4.2$ eV. This value compares well with previously experimentally and theoretically determined values, generally in the 2-10 eV range, for the deformation potentials in single-layer and multilayer MoS$_{2}$\cite{Cheiw12,Peel12,Jifeng12,Keliang13,Kaasbjerg12}.

\subsection{Carrier dynamics on short time scales}

Knowing $j(\omega)$ and $\Delta\sigma(\omega,u)$, we determine the photoexcited DC conductivity, $\Delta\sigma_{\rm DC}(u)$ (see Methods). The temporal resolution in our experiments is set by the measurement bandwidth of $\Delta\sigma(\omega,u)$,  $\sim$1.8 THz, and is approximately $(2\times1.8\ \textrm{THz})^{-1}=0.28$ ps.  Since the in-plane conductivity effective masses for electrons and holes are approximately the same in multilayer MoS$_{2}$\cite{Peel12}, one can write the DC conductivity as $\sigma_{\rm DC} = N e^{2}\tau_{\rm D}/m_{\rm c}^*$, where $N$ is the total mobile carrier density and equals the sum of the mobile electron density, $n$, and the mobile hole density, $p$. The change in the carrier density $\Delta N$ can be determined from the measurement of $\Delta \sigma_{\rm DC}$.

\begin{figure}[tbp]
	\centering
		\includegraphics[width=.47\textwidth]{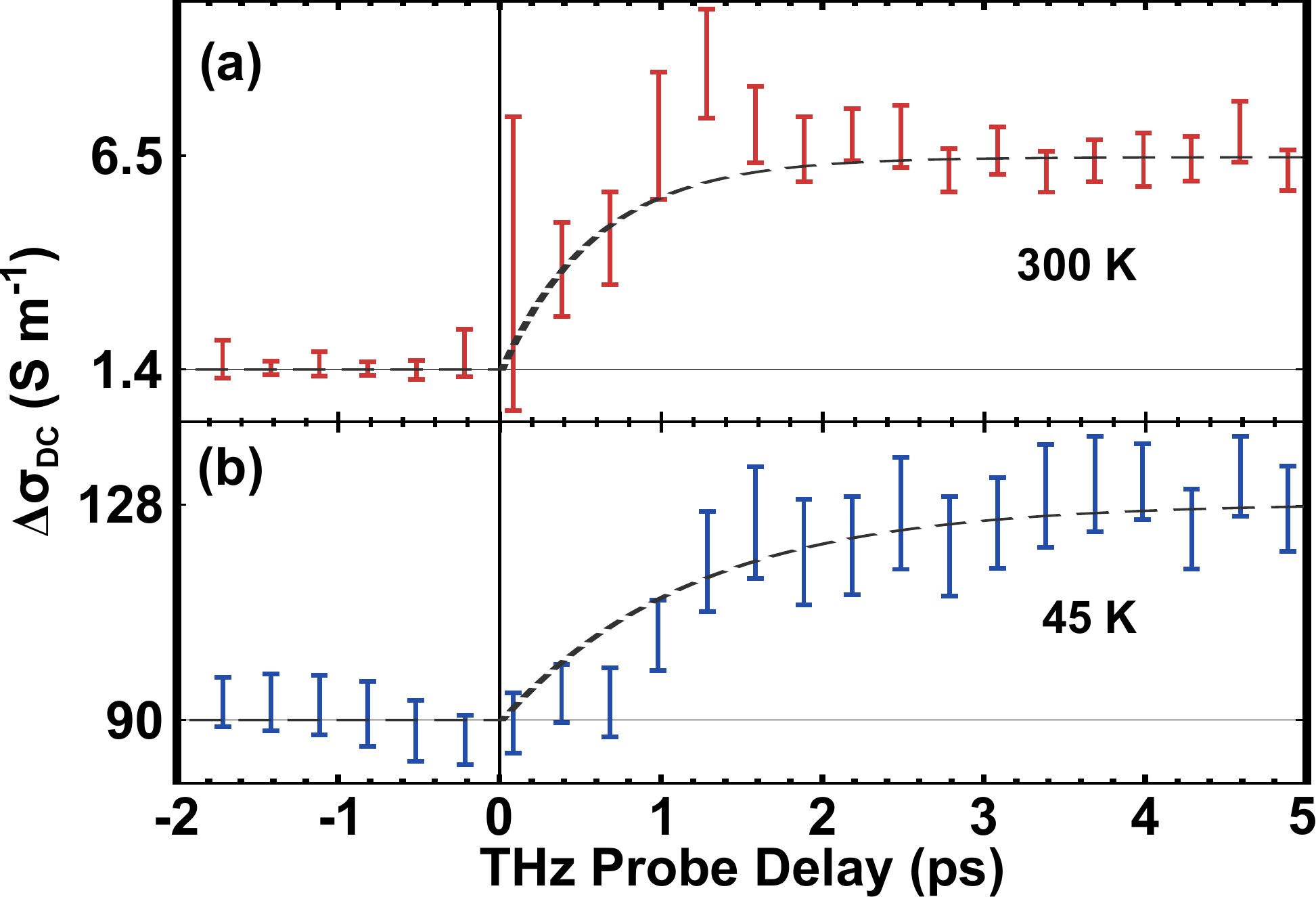}
	\caption{Intraband relaxation.  The change in DC conductivity, $\Delta \sigma_{\rm DC}(u)$, versus pump-probe delay on short time scales (-2 $\le u \le$ 5 ps).  We show data for two representative temperatures ((a) 300 K and (b) 45 K), each with a pump fluence of $\sim$1.2 $\mu$J cm$^{-2}$.  Immediately after the pump pulse, the conductivity increases, reaching its maximum value within 5 ps at all temperatures. The rise time is slower at 45 K than at 300 K.}
	\label{fig:TempDynamics}
\end{figure}

Figure \ref{fig:TempDynamics} shows the measured change in the conductivity $\Delta \sigma_{\rm DC}(u)$ on short time scales ($-2 \le u\le 5$ ps) for two different representative temperatures.  The pump fluence in each case is $\sim$1.2 $\mu$J/cm$^{2}$. Immediately after photoexcitation ($u=0$), the conductivity increases, reaching its maximum value within 5 ps at all temperatures.  The error-range displayed for each data point represents a 95\% confidence interval for the fitting parameter $\Delta \sigma_{\rm DC}(u)$. The dashed lines superimposed on the data represent fittings using exponential curves with 0.7 ps and 1.2 ps time constants at 300 K and 45 K, respectively.  Our data suggests that the conductivity reaches its peak value faster at higher temperatures.  As we discuss below, we attribute these short-time-scale dynamics to picosecond cooling of the electron and hole distributions.

As depicted in Figure \ref{fig:DeltaE}(b), the 1.58 eV pump photons excite electrons from the valence band maximum at the $\Gamma$ point to the conduction band minima at the $\Lambda$ point by a phonon-assisted (or impurity-assisted) indirect absorption process.  Since the indirect bandgap in multilayer MoS$_{2}$ is around 1.29 eV\cite{Mak10,Cheiw12}, the photoexcited electrons and holes are at an elevated temperature compared to the lattice immediately after photoexcitation and thermalization\cite{Jnawali13, Haining10}.  If the photon energy in excess of the indirect bandgap contributes to the kinetic energy of the carriers, we estimate the carrier temperature immediately after photoexcitation to be around 1100 K.  At this high temperature, the electron and hole distributions would be spread out in energy, with fast momentum scattering rates and reduced mobility due to optical phonons\cite{Kaasbjerg12,Lundstorm09}.  As the carrier distributions relax towards the band extrema and cool via optical phonon emission, the carrier momentum scattering rates decrease and the conductivity increases.  The rise in the conductivity following photoexcitation observed in our experiments is therefore indicative of carrier cooling.  The slower cooling observed at lower lattice temperatures can be attributed to a number of factors.  For example, the carrier cooling rates due to optical phonon emission are proportional to $n_{B}+1$, where $n_{B}$, the Bose occupation factor for phonons, is aproximately 20\% larger at 300 K compared to 45 K.  In addition, the heat capacity of optical phonons is around three orders of magnitude larger at 300 K compared to 45 K and, therefore, carrier cooling bottleneck due to the generation of hot phonons is much reduced at higher lattice temepratures\cite{Haining10}.    

\subsection{Carrier dynamics on long time scales}

\begin{figure}[tbp]
	\centering
		\includegraphics[width=.47\textwidth]{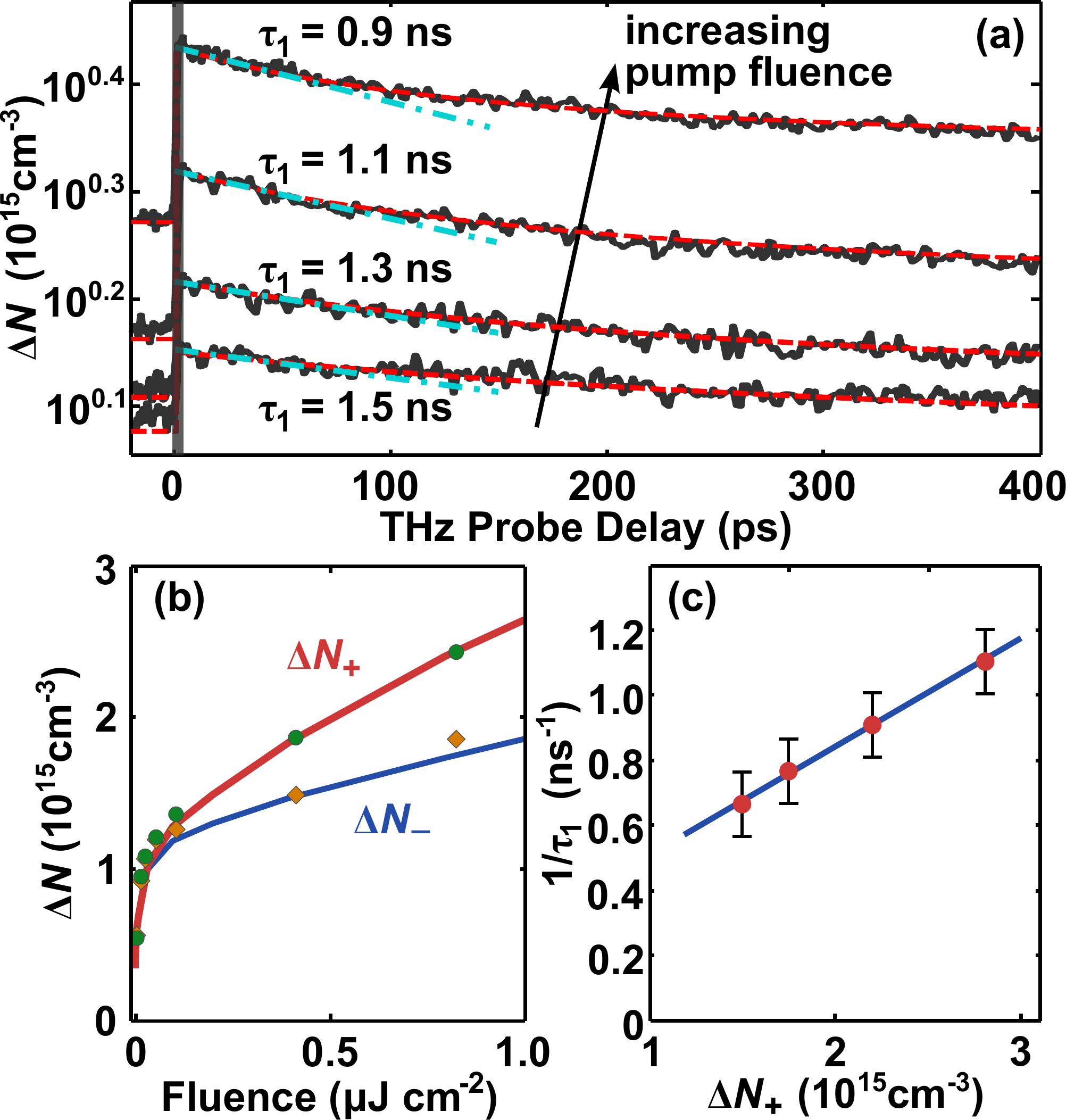}
	\caption{Density-dependent recombination.  (a) (Solid) The measured change in the carrier density $\Delta N(u)$ plotted on a log scale as a function of probe delay.  Shown are transients for four pump fluences (0.2, 0.4, 0.8, and 1.2 $\mu$J cm$^{-2}$) at 45 K.  (Blue dot-dash) Initial exponential relaxation corresponding to the estimated $\tau_{1}$.  (Red dash) Full transient simulations using Equations \ref{eq:model}.  (b) Measured $\Delta N_{+}=\Delta N(u=0^+)$ (circles) and $\Delta N_{-}=\Delta N(u=0^-)$ (diamonds) versus pump fluence at 45 K.  (Solid) Curves obtained from simulations of Equations \ref{eq:model}.  (c) The measured inverse decay time $1/\tau_{1}$ at 45 K versus $\Delta N_+$.  The slope of the line estimates the electron capture rate $An_{\rm d}$.}
	\label{fig:PowerDynamics}
\end{figure}
 
Figure \ref{fig:PowerDynamics}(a) shows the change in the mobile carrier density $\Delta N(u)$ at 45 K for various pump fluences, determined from measured $\Delta \sigma_{\rm DC}(u)$ on long time scales ($-20 \le u\le 400$ ps).  As expected, the carrier density decays after reaching a peak value in the first few picoseconds after photoexcitation.  We attribute this relaxation to carrier trapping and recombination by defects.  The carrier density decay rate is relatively fast in the first $\sim$100 ps after photoexcitation, after which the decay occurs on much longer time scales.  The maximum probe delay allowed by our setup, 400 ps, is unable to resolve these long time scales.  But the non-zero carrier densities observed at negative probe delays show that the photoexcited carrier density does not completely decay in the time interval, around 12.3 ns, between two successive optical pulses.  The carrier densities immediately before and after photoexcitation, as a function of pump fluence, reveal the carrier density dependence of the long term ($>$100 ps) recombination rates.  Figure \ref{fig:PowerDynamics}(b) shows the change in carrier density immediately before, $\Delta N_{-} = \Delta N(u=0^{-})$, and after, $\Delta N_{+} = \Delta N(u=0^{+})$, photoexcitation.  Both $\Delta N_{-}$ and $\Delta N_{+}$ increase with the pump fluence in a nonlinear fashion, so carrier-density-independent recombination times can be ruled out. We also estimate the initial recombination rate, $1/\tau_1$, versus pump fluence from the slope of the blue dash-dot lines in Figure \ref{fig:PowerDynamics}(a).  Figure \ref{fig:PowerDynamics}(c) shows that $1/\tau_{1}$ increases approximately linearly with $\Delta N_{+}$.  Any physical model that describes the observed carrier dynamics must account for the transition from the fast decay rates in the first $\sim$100 ps to the slow decay rates thereafter. In addition, the model must account for the severe nonlinearity of $\Delta N_-$ and $\Delta N_+$ versus pump fluence.

Here, we present a model of carrier relaxation based on trapping and recombination by optically-active defect states.  This model explains all features of the observed carrier dynamics at 45 K.  The natural MoS$_2$ used in our work is known to have defects such as vacancies, grain boundaries, and impurities\cite{Zhou13,Zhu14}.  Many of these defects are optically active in the near-IR wavelength region and appear in the optical absorption spectra\cite{Yuan14,Wang14,Wang13}.  We assume that the initial fast decay of the photoexcited carrier density is due to the capture of electrons by defect states in the bandgap\cite{Furchi14} (see Figure \ref{fig:DeltaE}(b)).  As these defect states become full of electrons, the decay rate decreases (due to Pauli blocking) and becomes limited by the capture of photoexcited holes from the valence band. The optical pump pulse excites electrons to the conduction band from both the valence band and also the defect states, and we assume that the defect states become empty immediately after the pump pulse\cite{Wang14,Wang13}. Our final assumption is that the defect states are assumed to be occupied by electrons in thermal equilibrium in our n-doped sample.

Ignoring thermal generation of electrons and holes from defect states, the rate equations for the decay of the electron and hole densities in our model are (see supplementary information),
\begin{eqnarray}
\frac{dn}{dt} & = & - A n_{\rm d} n^{2} (1-f_{\rm d})	\nonumber \\
\frac{dp}{dt} & = & - B n_{\rm d} n p f_{\rm d} 		\nonumber \\
\frac{df_{\rm d}}{dt} & = &  A n^2 (1-f_{\rm d}) - B n p f_{\rm d} \label{eq:model}
\end{eqnarray}
Here, $n$ and $p$ are the total mobile electron and hole densities, respectively, and $A$ ($B$) is the rate constant for electron capture (hole capture) by the defect states via the dominant Auger mechanisms in a n-doped semiconductor\cite{Landsberg92}.  The value $n$ is the sum of the equilibrium electron density, $n_0$, and the excess photoexcited electron density.  $n_{\rm d}$ is the density of defect states, and $f_{\rm d}$ is the electron occupation of the defect states, assumed to be unity in equilibrium.  Note that a phonon-assisted carrier trapping process would not have the carrier density dependence consistent with the observed decay rates\cite{Landsberg92}.  We varied the values of the fitting parameters $A$, $B$, $n_{\rm f}$, and $n_0$ in simulations to fit the data in Figures \ref{fig:PowerDynamics}(a) and \ref{fig:PowerDynamics}(b).  The simulations involved time-stepping the rate equations over many (1-5$\times10^{3}$) pump pulse cycles until a steady state was achieved.  The difference $\Delta N_+ - \Delta N_- \approx 0.52\times10^{15}$ cm$^{-3}$ per $\mu$J cm$^{-2}$ determined the mobile carrier density created by the pump pulse in our simulations.

The data in Figure \ref{fig:PowerDynamics} can inform the search for appropriate values of the fitting parameters.  Immediately after photoexcitation, since $n$ varies as $\Delta N_{+}/2$, the slope of the line in Figure \ref{fig:PowerDynamics}(c) estimates the product $A n_{\rm d}$.  Since the long term decay of the photoexcited density is limited by the capture of holes in defects, the values of $\Delta N_-$ estimate the product $B n_{\rm d}$.  Finally, the density of defect states governs how quickly they fill with electrons, therefore $n_{\rm d}$ is determined from the time at which the initial fast decay transitions to the slower decay, as measured in Figure \ref{fig:PowerDynamics}(a).  The values of fitting parameters that best fit the data are: $A=2.6\times10^{-21}$ cm$^6$ s$^{-1}$, $B=2.1\times10^{-23}$ cm$^6$ s$^{-1}$, $n_{\rm d}=5.0\times10^{14}$ cm$^{-3}$, and $n_0=2.7\times10^{14}$ cm$^{-3}$.  This value of $n_0$ corresponds to a DC conductivity of 12 S m$^{-1}$, which compares well with the value of conductivity obtained from THz transmission measurements of the unpumped sample, $\sigma_0=10-20$ S m$^{-1}$.  The model presented here closely agrees with the data, accurately reproducing the values of $\Delta N_-$, the relaxation curves, and the nonlinearity of $\Delta N$ versus pump fluence.

\section{Discussion}
In this paper, we presented measurements of the mobility of photoexcited carriers in multilayer MoS$_2$ using THz time-domain spectroscopy. The observed temperature dependence of the mobility for $T\le 200$ K indicates acoustic phonon scattering as the dominant mobility-limiting mechanism. The measured carrier momentum scattering rates are comparable to theoretical predictions based on phonon-scattering-limited transport\cite{Kaasbjerg12,Ma14}. In contrast, previously reported DC electrical measurements have suggested impurity or defect scattering as the dominant mobility-limiting mechanism in multilayer MoS$_2$\cite{Fivaz67,Kim12,Radisavljevic13,Bao13}.  Given the prevalence of defects and impurities in MoS$_2$\cite{Zhou13,Zhu14}, it is intriguing that we observe mobilities limited by acoustic phonon scattering.  The observed high mobilities at low temperatures in our experiments could be related to the fact that our measurements were performed at very high frequencies.  It is well known theoretically and experimentally that AC conductivity in disordered materials increases with the AC frequency\cite{Pollak61,Rockstad70,Pike72,Landauer52,Clerc90,Henning99}.  This phenomenon appears in the microscopic models of hopping conduction\cite{Pollak61,Rockstad70,Pike72} as well as in the classical models of the AC conductivity in inhomogeneous materials\cite{Landauer52,Clerc90,Henning99}. Since we did not see signatures of AC hopping conduction in our sample at any temperature, we believe models that describe the AC conductivity in inhomogeneous materials are more relevant to our observations\cite{Landauer52,Clerc90,Henning99}. Accordingly, we believe that the MoS$_{2}$ atomic layers in our sample consist of interspersed high and low mobility regions (due to grain boundaries, crystal defects, impurities, etc.\cite{Zhou13}) and the conductivity of the layers is determined by the resistive and capacitive couplings of these regions\cite{Landauer52,Clerc90,Henning99}. Even a small fraction of low mobility regions can significantly affect the DC conductivity in lower dimensions. The high mobility regions have a conductivity given by the Drude form. The AC conductivity of the sample is then expected to increase with the frequency until the low mobility regions are capacitively shorted out and the conductivity of the sample is then limited by the Drude conductivity of the high mobility regions. Although this model qualitatively explains the observation of the long momentum scattering times in our experiments, investigation of the frequency dependence of the sample conductivity in the low frequency region ($< 0.2$ THz) is needed to fully understand the nature of the transport and determine which model, if any, best describes the nature of the conduction.     
 
Carrier relaxation and recombination dynamics directly affect the performance of almost all electronic and optoelectronic devices. We have observed several time scales in the dynamics associated with carrier relaxation and recombination. The initial intraband relaxation (or carrier cooling) occurs within 5 ps. Carrier interband recombination appears to result from carrier trapping in optically active defect states, with recombination lasting over tens of nanoseconds due to slow hole capture in the defect states.  It is important to mention that simple bimolecular recombination (proportional to $n^2$) and direct interband Auger recombination (proportional to $n^3$) are not sufficiently nonlinear to account for the density dependence seen in Figure \ref{fig:PowerDynamics}(b).  Although our data shows that the recombination times in multilayer MoS$_{2}$ are long, they are short in comparison with other indirect bandgap semiconductors, such as high-quality Si or Ge, which have recombination times in excess of one microsecond at room temperature\cite{Gaubas06}.  We note here that our measurements might not have detected charge trapping dynamics occurring on much longer time scales ($\gg$10 ns) recently observed in MoS$_{2}$ photoconductive devices\cite{Cho14}.

The authors would like to acknowledge helpful discussions with Haining Wang, Michael G. Spencer, and Paul L. McEuen, as well as support from CCMR under NSF grant number DMR-1120296, AFOSR-MURI under grant number FA9550-09-1-0705, ONR under grant number N00014-12-1-0072, and the Cornell Center for Nanoscale Systems funded by NSF.

\section{Methods}
\subsection{Sample Preparation and Measurements}
The multilayer MoS$_2$ sample used in this study was cleaved from a large piece of natural MoS$_2$ (SPI Supplies).  We adhered the resulting flake to completely cover a 2 mm clear aperture and mounted the sample in a cryostat.  By measuring the broadband optical transmission interference fringes and using existing index of refraction data\cite{Evans65}, we determined that the average thickness of the flake was 4 $\mu$m, with variations of $\pm0.5$ $\mu$m across the aperture.  Electron-hole pairs were optically excited in the MoS$_2$ using sub-100 fs pulses from a Ti:Sapphire oscillator with a center frequency of 785 nm, pulse repetition rate of 81 MHz, and maximum fluence of 1.2 $\mu$J cm$^{-2}$.  We used synchronized, few-cycle THz pulses, generated and detected with photoconductive switches in a THz time-domain spectrometer\cite{Katzenellenbogen91}, to probe the excited carrier distribution.  Optical pump and THz probe beams were mechanically chopped at 400 and 333 Hz, respectively, and the detected photocurrent was demodulated with a lock-in amplifier at the sum frequency.

\subsection{Extraction of $\Delta \sigma(\omega,u)$ from Measurements}
The Fourier transform of Equation \ref{eq:J} is,
\begin{equation}
\Delta \tilde{J}(\omega,u) \approx E_0(\omega) \Delta \tilde{\sigma}(\omega,u)
\end{equation}
where, $\Delta\tilde{\sigma}(\omega,u)$ equals $\mathcal{F}[\Delta\sigma_{\rm DC}(u-t)j(t)]$. Here, $\mathcal{F}$ is the Fourier transform operator with respect to $t$. Measurement of  $\Delta \tilde{E}(t,u)$ enables one to obtain  $\Delta \tilde{J}(t,u)$ using Equation \ref{eq:DeltaJ}, and then $\Delta\tilde{\sigma}(\omega,u)$ can be obtained using the above Equation. In general, $\Delta\tilde{\sigma}(\omega,u)$ does not equal $\Delta\sigma(\omega,u) = \Delta \sigma_{\rm DC}(u) j(\omega)$\cite{Nienhuys05}. However, if the DC conductivity is changing slowly compared to the duration of the current impulse response $j(t)$ then $\Delta\tilde{\sigma}(\omega,u) \approx \Delta\sigma(\omega,u)$. Since the carrier density, and therefore the DC conductivity, varies slowly for $u\ge5$ ps in all of our measurements, we have extracted $\Delta\sigma(\omega,u)$ using this simple relation for $u\ge 5$ ps. But extracting $\Delta\sigma(\omega,u)$ from $\Delta\tilde{\sigma}(\omega,u)$ for cases when the DC conductivity varies rapidly, such as in Figure \ref{fig:TempDynamics}(b), requires two transformations\cite{Nienhuys05}:  First, $\Delta\tilde{\sigma}(t,u)=\mathcal{F}^{-1}[\Delta\tilde{\sigma}(\omega,u)]$, followed by, $\Delta\sigma(\omega,u)=\mathcal{F}[\Delta\tilde{\sigma}(t,u+t)]$. These transformations can add considerable noise, and we have employed them only to extract the DC conductivity changes for small pump-probe delays, $-2<u<5$ ps, and low temperatures.

\onecolumngrid
\newpage

\section{Supplementary Information for ``High Intrinsic Mobility and Ultrafast Carrier Dynamics in Multilayer Metal Dichalcogenide MoS$_2$''}

\subsection{Expressions for the transmission of THz pulses}
\begin{figure}[tbp]
	\centering
		\includegraphics[width=.45\textwidth]{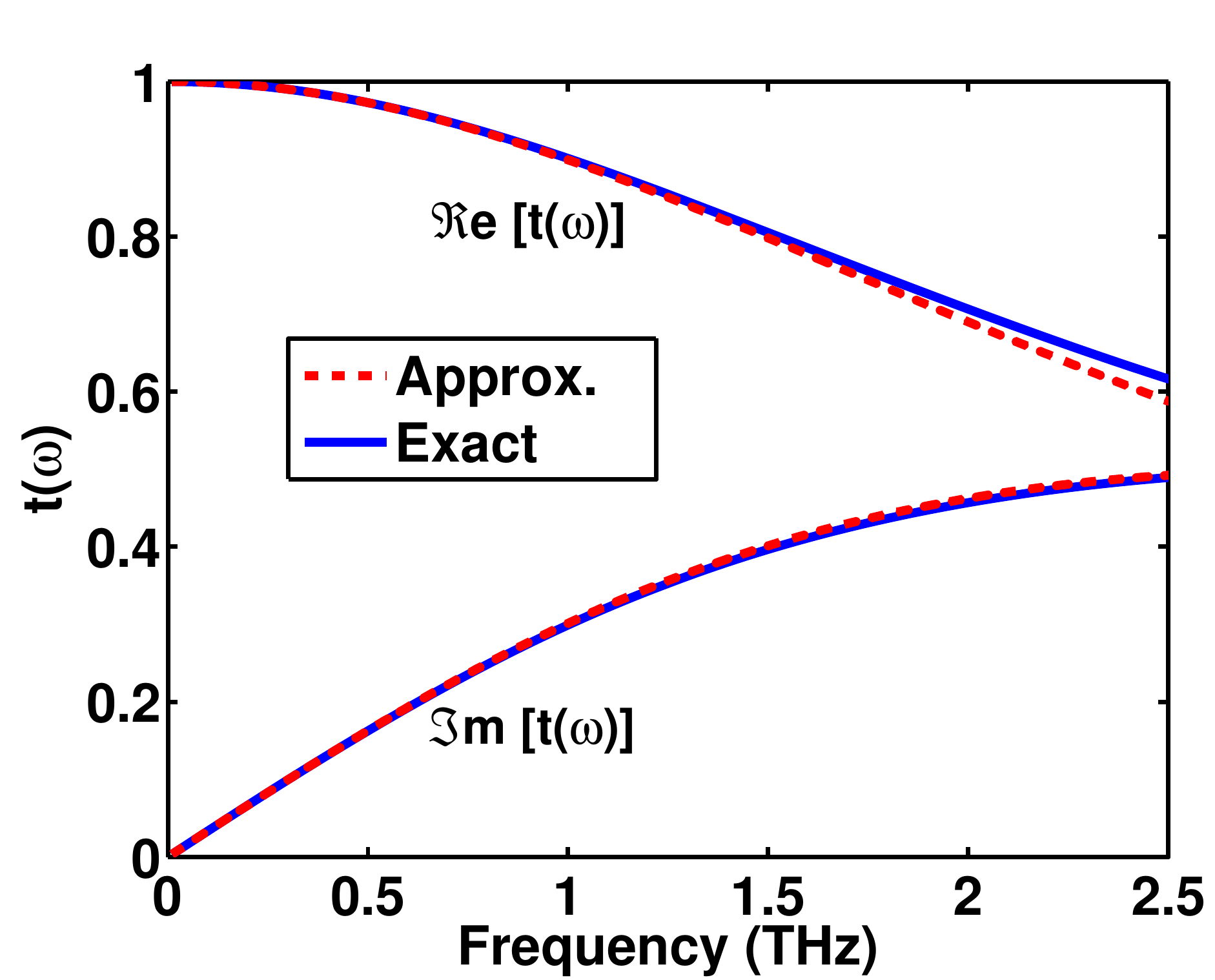}
	\caption{The real and imaginary parts of the transmission are shown for the case of the exact expression and the approximate expression for $n(\omega)=3$ and $d=4$ $\mu$m.}
  	\label{fig:trans}
\end{figure}

We use Equation 1 of the main text to estimate the frequency-dependent transmission of THz pulses through the MoS$_2$ sample.  Equation 1 can be derived as follows. We start with the exact solution for the transmission of light through an etalon with a refractive index $n(\omega)$ and thickness $d$, surrounded by air ($n_{\rm air}=1$),  
\begin{equation}
t(\omega)=\frac{E_t(\omega)}{E_i(\omega)}=\frac{4 n(\omega) e^{i\frac{\omega d}{c}(n(\omega)-1)}}{(n(\omega)+1)^2-(n(\omega)-1)^2e^{-2i\frac{\omega d}{c}n(\omega)}}
\end{equation}
In the limit, $\omega d/c \ll 1$,
\begin{equation}
t(\omega) \approx \left[1-i\frac{\omega d}{2c}(n(\omega)^2-1)\right]^{-1} \label{eq:approx}
\end{equation}
For the measured THz refractive index of MoS$_{2}$, $n(\omega) \approx 3.0$, and the thickness of our sample, $d \approx 4$ $\mu$m, the approximation in the above expression is extremely good, as illustrated in Figure \ref{fig:trans}. 
In the case of a conductive medium, with conductivity $\sigma(\omega)$, the complex refractive index is, $n^{2}(\omega)=n^2+i\sigma(\omega)/\omega\epsilon_o$. Here, $n$ is the frequency-independent refractive index and $\epsilon_o$ is the permittivity of free space.  The inverse Fourier transform of the relation in Equation \ref{eq:approx} gives,
\begin{equation}
\frac{\partial E_{t}(t)}{\partial t} = -\frac{2}{\epsilon_0(n^{2}-1)\eta_0d}\left[ E_{t}(t) - E_{i}(t) \right] - \frac{1}{\epsilon_0(n^{2}-1)} J(t)
\end{equation}
Subtracting versions of the above equation in the presence and absence of optical pumping one obtains,
\begin{equation}
\Delta J(t,u) = - \frac{2}{\eta_0 d}\Delta E(t, u) - \epsilon_0 (n^2-1)\frac{\partial\Delta E(t, u)}{\partial t} \label{eq:DeltaJS}
\end{equation}
Equation 2 in the text follows directly from the above relation following the substitution $u\to u-t$.

\subsection{Derivation of Equation 3 in the Main Text}
\begin{figure}[tbp]
	\centering
		\includegraphics[width=.45\textwidth]{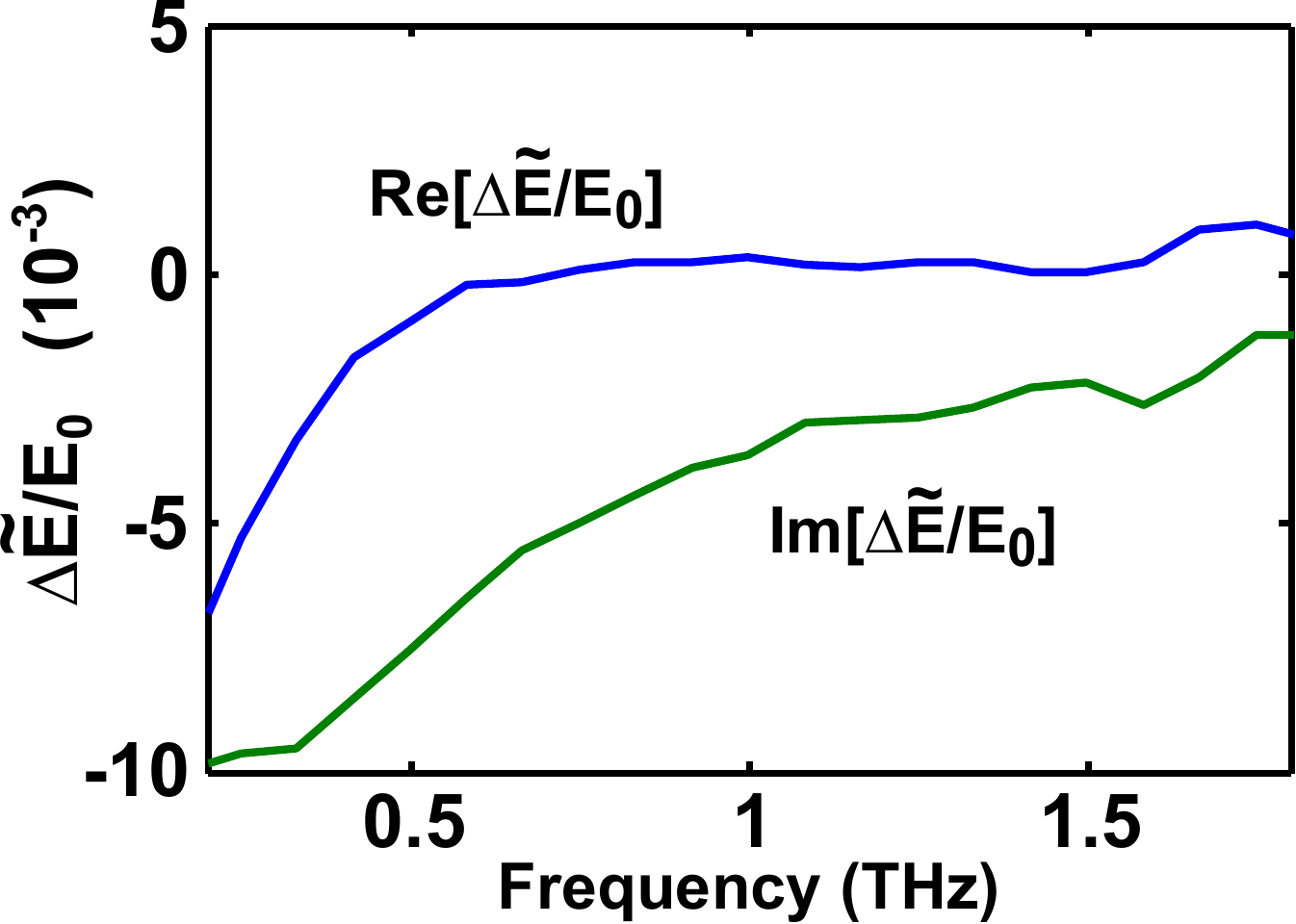}
	\caption{The real and imaginary parts of the ratio $\Delta \tilde{E}(\omega,u)/E_0(\omega)$ for $u=5$ ps and maximum pump fluence at 45 K.  For all other temperatures, fluences, and pump-probe delays, $\Delta \tilde{E}(\omega,u)/E_0(\omega)$ is either similar or smaller.}
  	\label{fig:ratio}
\end{figure}

In this Section, we wish to establish Equations 3 and 6 of the main text. We assume that the current density $J(t,u)$ in the sample for pump-probe delay $u$ in the presence of the field $E(t,u)$ of a THz pulse and a time-dependent DC conductivity $\sigma_{\rm DC}(t+u)$ is\cite{Nienhuys05},
\begin{equation}
J(t,u) = [ E(t,u)\sigma_{\rm DC}(t+u)] \otimes j(t) \label{eq:JS}
\end{equation}
Here, $j(t)$ is the normalized current impulse response ($\int_{-\infty}^{\infty} dt j(t)=1$) and $\otimes$ is the convolution operator with respect to $t$. This is the most general way of writing the current response if the current density obeys the operator equation,
\begin{equation}
 \hat{L} J(t,u) \propto E(t,u)\sigma_{\rm DC}(t+u)
\end{equation}
where, $\hat{L}$ is a linear-time-invariant (LTI) differential operator. Equation \ref{eq:JS} gives,
\begin{equation}
\tilde{J}(t,u) = J(t,u-t) = E(t,u-t) \otimes [ \sigma_{\rm DC}(u-t) j(t)] = \tilde{E}(t,u) \otimes \tilde{\sigma}(t,u)
\end{equation}
The change $\Delta \tilde{J}(t,u)$ in the current density in the presence and absence of pumping is $\tilde{J}_{\rm p}(t,u) - J_0(t)$. Here, the subscripts p and 0 indicate the value with and without optical pumping, respectively. Therefore, one obtains,
\begin{align}
	\Delta\tilde{J}(t,u) & = \tilde{\sigma}_{\rm p}(t,u) \otimes \tilde{E}_{\rm p}(t,u) - \sigma_0(t) \otimes E_0(t) \nonumber \\
	& = (\Delta\tilde{\sigma}(t,u)+\sigma_0(t))\otimes(\Delta\tilde{E}(t,u)+E_0(t))-\sigma_0(t)\otimes E_0(t) \nonumber \\
	& = \Delta\tilde{\sigma}(t,u)\otimes E_0(t) + \sigma_0(t)\otimes \Delta\tilde{E}(t,u) + \Delta\tilde{\sigma}(t,u)\otimes\Delta\tilde{E}(t,u)
\end{align}
By applying a Fourier transform with respect to $t$ to each term, we have,
\begin{equation}
	\Delta\tilde{J}(\omega,u) = \Delta\tilde{\sigma}(\omega,u)E_0(\omega) + \sigma_0(\omega)\Delta\tilde{E}(\omega,u) + \Delta\tilde{\sigma}(\omega,u)\Delta\tilde{E}(\omega,u)
	\label{eq:dJ}
\end{equation}
Equations 3 and 6 of the main text will be valid provided the final two terms on the right hand side of Equation \ref{eq:dJ} are negligibly small compared to the first term. In Figure \ref{fig:ratio}, we show a representative measured ratio of $\Delta\tilde{E}(\omega,u)/E_0(\omega)$, which is at most on the order of $10^{-2}$ for the highest pump fluence used in our experiments. At all other temperatures, pump fluences, and pump-probe delays, this ratio is either similar or smaller.  Since $\Delta\tilde{E}(\omega,u)\ll E_0(\omega)$ at all frequencies, we can safely neglect $\Delta\tilde{\sigma}(\omega,u)\Delta\tilde{E}(\omega,u)$ compared to $\Delta\tilde{\sigma}(\omega,u)E_0(\omega)$ in Equation \ref{eq:dJ}. Next, using Equation 2 from the main text, we see that the ratio $|\Delta\tilde{J}(\omega,u)/\Delta\tilde{E}(\omega,u)| \ge 2/(\eta_0d) \approx 1300$ S m$^{-1}$.  In contrast, the measured value of the unpumped DC conductivity $\sigma_0(\omega = 0)$ is smaller than $20$ S m$^{-1}$ and, therefore, $|\sigma_0(\omega)| \ll |\Delta\tilde{J}(\omega,u)/\Delta\tilde{E}(\omega,u)|$ at all frequencies. Consequently, we can also neglect $\sigma_0(\omega)\Delta\tilde{E}(\omega,u)$ compared to $\Delta\tilde{\sigma}(\omega,u)E_0(\omega)$ in Equation \ref{eq:dJ}. In this way, we empirically verify that our use of the approximations in Equations 3 and 6 in the main text is valid for our our experiments. Theoretically, it can be shown that,
\begin{equation}
\frac{\sigma_0(\omega)\Delta\tilde{E}(\omega,u)}{ \Delta\tilde{\sigma}(\omega,u)E_0(\omega)} \approx - \frac{\displaystyle \frac{\eta_0\sigma_{0}(\omega) d}{2}}{\displaystyle 1-i\frac{\omega d}{2c} (n^{2}-1) + \frac{\eta_0\sigma_0(\omega) d}{2} }  
\end{equation}
The second term in Equation \ref{eq:dJ} can be neglected provided $\eta_0|\sigma_0(\omega)| d/2 \ll 1$, which is indeed the case in our experiments.  

\subsection{Model for electron-hole recombination via mid-gap defect states}
\begin{figure}[tbp]
	\centering
		\includegraphics[width=.45\textwidth]{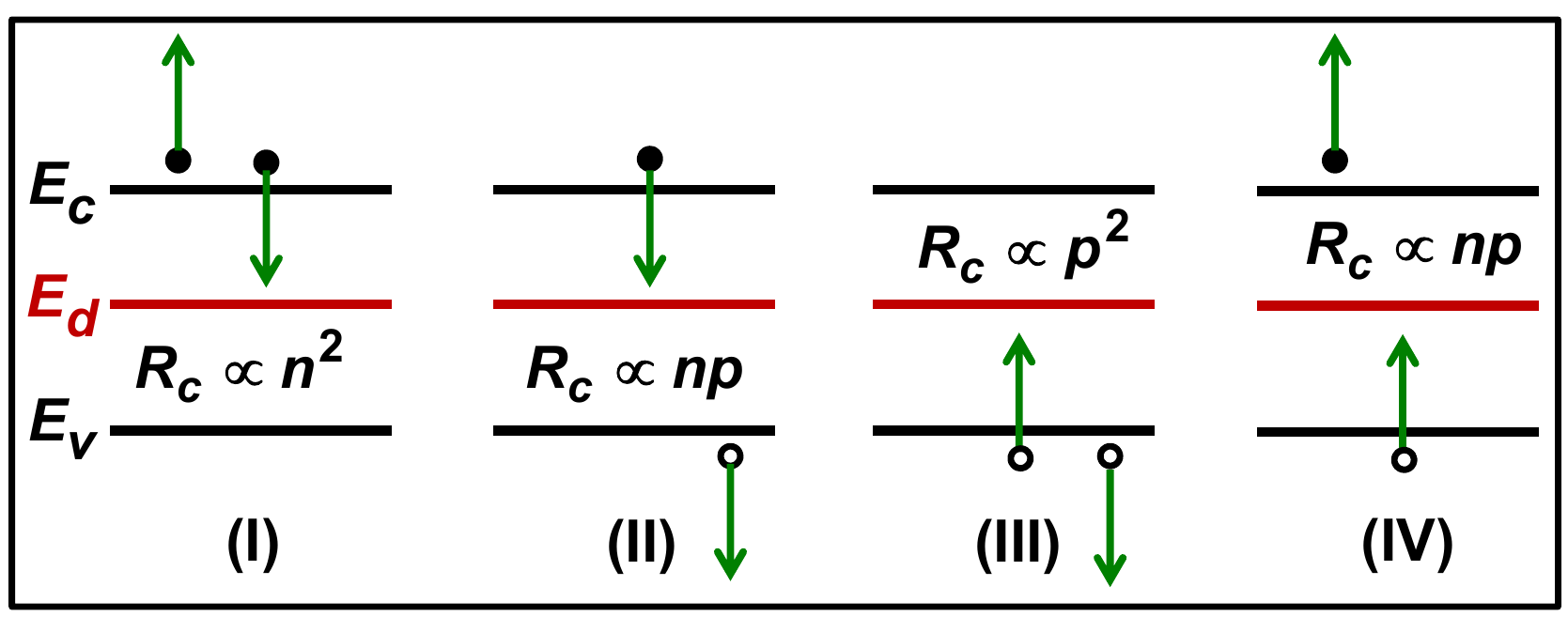}
	\caption{The four basic Auger processes for the capture of electrons and holes at defect states are illustrated. The energies of the conduction band bottom, valence band top, and the defect levels are $E_{\rm c}$, $E_{\rm v}$, and $E_{rm d}$, respectively. In each case, the approximate carrier density dependence of the capture rates are indicated\cite{Landsberg92}.}
  	\label{fig:auger}
\end{figure}
There are essentially two main mechanisms for the capture/emission of electrons and holes at/from localized crystal defects\cite{Landsberg92}: (1) Phonon-assisted processes, and (2) Auger processes. Phonon-assisted processes can be single-phonon processes or multi-phonon processes, including phonon-cascade processes\cite{Ridley13}. In phonon-assisted processes the capture rates (units: cm$^{-3}$s$^{-1}$), tend to go linearly with the carrier density (i.e. the capture times are independent of the carrier density). The capture times observed in our experiments are carrier density dependent (inverse electron capture times increase linearly with the electron density), as shown in Figure 5(c) of the main text. The capture times in Auger processes are carrier density dependent. Figure \ref{fig:auger} shows the four basic Auger processes for the capture of electrons ((I) and (II)) and holes ((III) and (IV)) at defects. The corresponding emission processes are the just the inverse of the capture processes. The rate equations for each Auger process (and its inverse) can be written using Figure \ref{fig:auger}. For example, the electron density rate equation for process (I) and its inverse is,
\begin{equation}  
\frac{dn}{dt} =  - A n_{\rm d} n^{2} (1-f_{\rm d}) + A  n_{\rm d} n n^{\ast} f_{\rm d} 
\end{equation}
Here, $n$ is the electron density, $A$ is the rate constant for electron capture by the defect state, $n_{d}$ is the defect density, and $f_{d}$ is the occupation of the defect state. The first terms describes the capture process and the second term describes the emission process. The value of the constant $n^{\ast}$ can be determined by using the fact that in thermal equilibrium $dn/dt=0$:
\begin{equation}
n^\ast = n_{0}\frac{1-f_{\rm do}}{f_{\rm do}}
\end{equation}
where $n_{0}$ is the equilibrium electron density and $f_{\rm do}$ is the equilibrium defect occupation. If in equilibrium $f_{\rm do}\approx 1$, as is expected for defects deeper than a few $k_{\rm B}T$ in an n-doped material, then $n^\ast$ can be assumed to be negligibly small and electron generation from the defect states can be ignored in the above equation. Process (I) and process (II) can have comparable magnitudes\cite{Landsberg92}. So, ignoring emission processes, the rate equation for the electron density becomes,
\begin{equation}  
\frac{dn}{dt} =  - A n_{\rm d} n^{2} (1-f_{\rm d}) - C n_{\rm d} np (1-f_{\rm d})
\end{equation}
$A$ and $C$ are the rate constants for electron capture by the defect state corresponding to processes (I) and (II) in Figure \ref{fig:auger}, respectively. In our experiments, since our MoS$_{2}$ sample is n-doped, the first term of the right hand side is more important for small pump fluence values (when the hole density is small) compared to the second term. For large pump fluence values, both the electron and the hole densities can become comparable and the second term on the right hand side may not be ignored. However, at large pump fluences the effect of the second term is indistinguishable from the first term in our pump-probe experiments (since both terms would result in the inverse electron capture time to increase linearly with the photoexcited carrier density). Since the pump fluences used in our experiments are relatively small (and the maximum photoexcited carrier density is in the low $10^{15}$ cm$^{-3}$), we have chosen to ignore the second term, corresponding to process (II), in the above Equation for simplicity. Similarly, ignoring emission processes, the rate equation for the hole density becomes,  
\begin{equation}  
\frac{dp}{dt}  =  - B n_{\rm d} n p f_{\rm d} - D n_{\rm d} p^{2} f_{\rm d}
\end{equation}
Here, $p$ is the hole density. $D$ and $B$ are the rate constants for hole capture by the defect state corresponding to processes (III) and (IV) in Figure \ref{fig:auger}, respectively. Again, since our MoS$_{2}$ sample is n-doped and the pump fluences used in our experiments are small, we have chosen to ignore the second term, corresponding to process (IV), in the above Equation for simplicity.

\end{document}